\documentstyle[12pt,aaspp]{article}
\newcommand{\fulljustify}{\rightskip=0pt}
\fulljustify     

\newcommand{\beq}{\begin{equation}}              
\newcommand{\eeq}{\end{equation}}             
\newcommand{\beqa}{\begin{eqnarray}}              
\newcommand{\eeqa}{\end{eqnarray}}             

\def\mg{$\mu$G~}
\def\myref{\bibitem{dummy}\vspace{-2pt}}
\let\gsim=\gtrsim
\let\lsim=\lesssim
\begin{document}
\title{Faraday Rotation of Microwave Background Polarization\\
by a Primordial Magnetic Field}
\author{Arthur Kosowsky${}^{1,2}$ and Abraham Loeb${}^1$}
\affil{${}^1$Center for Astrophysics, Harvard University,
60 Garden St., Cambridge MA 02138}
\affil{${}^2$Department of Physics, Lyman Laboratory, Harvard University,
Cambridge, MA 02138}
\begin{abstract}
\fulljustify
The existence of a primordial magnetic field 
at the last scattering surface may
induce a measurable Faraday rotation 
in the polarization of the cosmic microwave background. 
We calculate the magnitude of this effect by
evolving the radiative transfer equations for the
microwave background polarization through the
epoch of last scatter, in the presence of a 
magnetic field.
For a primordial field amplitude 
corresponding to a present value of $10^{-9}{\rm G}$
(which would account for the observed galactic field
if it were frozen in the pre-galactic plasma),
we find a rotation
angle of around $1^\circ$ at a frequency of 30 GHz. 
The statistical detection
of this signal is feasible with future maps of the
microwave background.

\bigskip
\noindent
\end{abstract}
\keywords{magnetic fields--cosmology:theory--cosmic microwave background}
\bigskip
\bigskip
\centerline{Submitted to {\it The Astrophysical Journal}, Jan. 1996} 

\clearpage
\section{Introduction}

The origin of the observed \mg magnetic fields in spiral
galaxies has been a long-standing puzzle
for more than three decades (Rees 1987; Kronberg 1994). On the one hand,
these magnetic fields
could have resulted from an
exponential dynamo amplification of
a small seed field
with an $e$-fold period of a galactic rotation (Parker 1979;
Zel'dovich, Ruzmaikin, \& Sokoloff 1983; Field 1994).
Alternatively, these fields may have already been
frozen in the primordial plasma before
galaxies formed  (Hoyle 1958; Piddington 1964, 1972; Ohki et al. 1964),
and could have even affected the evolution of structure
in the universe (Wasserman 1978; Kim, Olinto, \& Rosner 1994).
The origin of the primordial
field is still a subject of speculation (see, e.g. 
Turner \& Widrow 1988; Quashnock, Loeb, \& Spergel 1989; 
Vachaspati 1991; Ratra 1992).

In the past, various indirect theoretical arguments were used 
to favor the dynamo amplification 
mechanism over the primordial origin 
alternative (Zel'dovich, Ruzmaikin, \& Sokoloff 1983).
However, recent theoretical studies argue
that a galactic dynamo should saturate due to the rapid
growth of a fluctuating  small-scale
field before it can actually result in
a coherent large-scale field  
of the type observed in galactic disks  
(Kulsrud \& Anderson 1992; Vainshtein \& Cattaneo 1992;
Vainshtein, Parker, \& Rosner 1993; Cattaneo 1994). 
The view 
that the galactic field 
may, in fact, be primordial gains additional
support from observations of
damped Ly$\alpha$ absorption systems in QSO spectra
at $z_{\rm abs}\sim 2$.
These systems, which are thought to be the progenitors of
galactic disks (Lanzetta et al. 1995), add
Faraday rotation to their background QSOs, 
consistent with them having \mg 
fields  (Welter, Perry, \& Kronberg 1984;
Wolfe, Lanzetta, \& Oren 1992; Kronberg, Perry, \& Zukowski 1992).
Since these systems exist at an epoch when
their rotation period is not exceedingly small
compared to their age, this result could pose
a problem to the standard dynamo hypothesis.
However, the current data may not be sufficient 
to draw any firm statistical conclusions about absorption systems
beyond a redshift of 0.4 (Perry, Watson, \& Kronberg 1993).

The potential existence of a primordial 
magnetic field is also consistent with observations
of clusters of galaxies. Faraday rotation measurements of radio
sources inside and behind clusters 
indicate strong magnetic fields in many
of them (Kim et al. 1990, 1991; 
Taylor \& Perley 1993).
The detected cluster fields have a typical magnitude of a few
$\mu$G and a coherence length of $10^{1-2}$ kpc.
The cores of several clusters contain tangled
magnetic fields with amplitudes
as high as $\sim 10^{1-2}{\mu}$G 
(Dreher et al. 1987; Perley \& Taylor 1991; Taylor \& Perley 1993; 
Ge \& Owen 1993). In the outer halos of clusters,
lower limits $\gsim 0.1\mu$G were set on the field amplitude,
by combining measurements of synchrotron radio-emission
from relativistic electrons in these halos 
together with lower limits on the associated hard X-ray emission 
due to Comptonization of the microwave background 
(Rephaeli 1979; Rephaeli et al. 1987; Rephaeli, \& Gruber 1988).

If primordial in origin, the \mg galactic field could 
have resulted
directly from the adiabatic compression  
of a cosmological field, $B_0\sim (10^{-10}$--$10^{-9})$G.
%
Using a sample of 309 galaxies and quasars with a small
intrinsic rotation measure, Vallee (1990)
was able to set an upper limit of $10^{-9}{\rm G}\times
(\Omega_{\rm IGM}h/0.01)^{-1}$ on the magnitude 
of a cosmological magnetic field which is
coherent on the scale of the horizon; 
here, $\Omega_{\rm IGM}$ is the ratio between 
the ionized gas density
in the intergalactic medium and the critical density,
and $h$ is the Hubble constant $H_0$
in units of $100~{\rm km~s^{-1}~Mpc^{-1}}$.
However, this limit is
based on Faraday--rotation measures,
for which the contributions of 
field reversals along the line of sight
average-out in a tangled field configuration. The 
limit is weakened 
to a value $\gsim 3\times 10^{-8} {\rm G}\times
(\Omega_{\rm IGM}h/0.01)^{-1}$
if the cosmic field is coherent only on scales 
$\lsim 10 {\rm Mpc}$ (see also Kronberg 1994). 

In this paper, we propose a direct empirical probe of 
primordial magnetic fields.
It is by now established that the microwave background 
should have acquired a measurable level of
polarization at decoupling 
(see, e.g. Kosowsky 1996; Coulson, Crittenden, \& Turok 1995).
If a primordial magnetic field is present 
at the last scattering surface of the microwave photons, it will
cause Faraday rotation of the direction of
linear polarization.
Since the rotation angle depends on wavelength,
it is possible to infer this effect by comparing 
the polarization vector of the microwave sky in a given direction
at two different frequencies. Since both the magnetic field
amplitude ($\propto [1+z]^2$)
and the baryonic density ($\propto [1+z]^3$)
increase rapidly with redshift, this effect
could potentially have measurable consequences
at the high redshift of decoupling.

We can roughly estimate the expected rotation angle of the microwave
background polarization as follows. 
Monochromatic radiation of frequency $\nu$ passing through a
plasma in the presence of a magnetic field ${\bf B}$ along the propagation
direction $\bf\hat q$ 
will have its linear polarization vector rotated at the rate
\begin{equation}
{d\varphi\over dt} = {e^3 x_e n_e\over 2\pi m^2\nu^2} ({\bf B\cdot \hat q}),
\label{faraday}
\end{equation}
where $e$ and $m$ are the
electron charge and mass, $n_e$ is the total
number density of electrons, and $x_e$ is the ionization fraction.
(Throughout the paper we use natural units with $\hbar=c=G=1$.)
The optical depth for scattering 
is of order unity out to the redshift of decoupling when
polarization is generated, i.e. we may substitute
$\int x_e n_e dt\approx 1/\sigma_T$ where
$\sigma_T$ is the Thomson cross-section. 
The rms rotation angle can then be easily estimated 
from the time--integral of
equation~(\ref{faraday}) by noticing that $B/\nu^2$ is time--independent
and by averaging $\varphi^2\propto ({\bf B\cdot \hat q})^2$ over all 
possible orientations of ${\bf B}$,
\begin{equation}
\langle\varphi^2\rangle^{1/2}\approx 
{e^3 B_0\over 2{\sqrt{2}}\pi m^2 \sigma_T \nu_0^2}
= 1.6^\circ 
\left({B_0\over 10^{-9}{\rm Gauss}}\right)
\left({30~{\rm GHz}\over \nu_0}\right)^{2} ,
\label{estimate}
\end{equation}
where $B_0$ is the current amplitude of the cosmological magnetic field,
and $\nu_0$ is
the observed frequency of the radiation. Note that equation~(\ref{estimate})
is independent of cosmological parameters.
For a primordial field of $10^{-9}~{\rm G}$ which
could result in the observed galactic field, 
we therefore expect a rotation measure of order 
$1.6~{\rm deg~cm^{-2}}=280~{\rm rad~m^{-2}}$.
This rotation is considerable by astrophysical standards and could
in principle be measured. The exact value
of the rotation measure is, however, sensitive to the growth history
of the microwave background polarization
through the surface of last scatter.
A larger Faraday rotation is expected in anisotropic cosmological
models, previously investigated by Milaneschi \& Fabbri (1985).

In this paper, we perform a detailed calculation of the above Faraday
rotation signal.  We work in the context of inflation-type models with
adiabatic initial fluctuations, and assume no early reionization.  The
rotation generated at the surface of last scatter depends only on the
ionization history and is insensitive to most cosmological parameters.
Section 2 develops the formalism for calculating the microwave background
polarization including Faraday rotation.  In Section 3 we present numerical
results, including the dependence of the rotation angle on the mean mass
density and baryon density of the universe.  Finally, Section 4 discusses 
the prospects for detecting this effect.

\section{Formalism}

The evolution of the cosmic microwave background is described
by a set of radiative transfer equations for the Fourier modes
of the radiation brightnesses $\Delta_I({\bf k},{\bf\hat q},\eta)$,
$\Delta_Q({\bf k},{\bf\hat q},\eta)$, 
and $\Delta_U({\bf k},{\bf\hat q},\eta)$,
where the subscripts refer to the standard Stokes parameters, the
Fourier mode wavevector 
is given by $\bf k$, the radiation propagation direction
is given by the unit vector $\bf\hat q$, and
$\eta=\int (1+z)dt$ is conformal time. 
For pure blackbody fluctuations, the temperature deviation is
$\Delta T/T_0$ = $\Delta_I/4$, with $T_0=2.726\pm0.010$ K 
(Mather et al. 1993) the mean temperature.
In the case of no magnetic
fields, the transport equations depend only on 
$\mu = \cos({\bf\hat k\cdot\hat q})$, the cosine of the angle between a given
Fourier mode and the propagation direction. Also, the isotropy
of the Universe implies that the equations depend only on
$k = |{\bf k}|$ and not on the direction of the Fourier component.
In the following
calculation, we assume a uniform magnetic field and
extract mean results by averaging over the entire sky; then the
equations still depend only on $k$ and $\mu$ and not on $\bf\hat q$
and $\bf k$ separately. This is equivalent to assuming that the
magnetic field is coherent on the scale of the width of the
last scattering surface, a comoving scale of about 5 Mpc. This 
assumption is natural if the coherent magnetic field observed in galaxies
came from a primordial origin, since galaxies 
were assembled from a comoving scale of a few Mpc.

Including the Faraday mixing term, the radiative transport
equations in comoving coordinates
are (Bond \& Efstathiou 1984; Kosowsky 1996)
\begin{eqnarray}
\dot\Delta_I + ik\mu(\Delta_I - 4\Phi) &=& -4\dot\Psi
-\dot\tau \left[
\Delta_I - \Delta_{I\,0} + 4 v_b\mu 
- {1\over 2} P_2(\mu)(\Delta_{I\,2} + \Delta_{Q\,2} - \Delta_{Q\,0}) 
\right]\cr
\dot\Delta_Q + ik\mu\Delta_Q &=& 
-\dot\tau \left[ \Delta_Q  
+ {1\over 2}\left(1-P_2(\mu)\right)
(\Delta_{I\,2} + \Delta_{Q\,2} - \Delta_{Q\,0}) \right]
+ 2\omega_{\bf B}\Delta_U \cr
\dot\Delta_U + ik\mu\Delta_U &=& -\dot\tau\Delta_U 
- 2\omega_{\bf B}\Delta_Q ,
\label{radtrans}
\end{eqnarray}
where $\Psi$ and $\Phi$ are scalar metric perturbations in
the Newtonian gauge, $v_b$ is the baryon velocity, and
$\dot\tau$ is the differential optical depth, defined by
$\dot\tau = x_e n_e \sigma_T a/a_0$ with $x_e$ the ionization
fraction, $n_e$ the total electron density, $\sigma_T$ the
Thomson cross-section, and $a$ the scale factor normalized
to $a_0$ today. Dots over quantities represent derivatives
with respect to conformal time $\eta$. 
The numerical subscripts on the radiation brightnesses
indicate moments defined by an expansion of the
directional dependence in Legendre polynomials $P_\ell(\mu)$:
\begin{equation}
\Delta_{I\,\ell}(k) \equiv {1\over 2}\int_{-1}^1 d\mu\,
P_\ell(\mu) \Delta_I(k,\mu).
\label{moments}
\end{equation}
A detailed derivation of equations~(\ref{radtrans}) can be found
in Kosowsky (1996).

The mixing terms between $\Delta_Q$ and $\Delta_U$ account
for the effect of Faraday rotation; these
terms follow directly from the definition of the Stokes parameters.
The conformal Faraday rotation rate is given by 
\begin{equation}
\omega_{\bf B}\equiv {d\varphi\over d\eta} = {d\varphi\over dt}{a\over a_0} ,
\label{rate}
\end{equation}
where $\varphi$ is the rotation angle of the polarization vector. 
For a given magnetic field and direction of photon propagation,
the rotation rate $d\varphi/dt$ is given by equation~(\ref{faraday}).
The time dependence of the gravitational potentials $\Phi$ and $\Psi$
and the ionization fraction $x_e$ then completely determine the
evolution of the radiation brightnesses.

Hu and Sugiyama (1995a,b) have demonstrated how to solve
the radiative transport equations semi-analytically,
using analytic fits to the evolution of the potentials
and the ionization fraction. This
formalism has recently been extended to include polarization
(Zaldarriaga \& Harari 1995). Essentially, the tight-coupling
solution describing the primordial plasma is modified by a
damping factor to account for diffusion damping through
recombination, and the resulting photon distribution free-streams
to the present epoch. We use this approach to give the temperature
fluctuations, which then source the polarization fluctuations.
Here we present the modifications to the tight-coupling
solution necessary to incorporate Faraday rotation; for other
details of the calculation, see Hu \& Sugiyama (1995a,b). 

The polarization brightness $\Delta_Q$ is sourced by the
function $S_p \equiv \Delta_{I\,2} + \Delta_{Q\,2} - \Delta_{Q\,0}$,
and $\Delta_U$ is generated as $\Delta_Q$ and $\Delta_U$ are
rotated into each other. In the absence of magnetic
fields, $\Delta_U$ retains its tight-coupling value of zero.
Expanding to second order in the tight-coupling parameter 
$\dot\tau^{-1}$ gives the evolution equation
(Zaldarriaga \& Harari 1995)
\begin{equation}
\dot S_p + {3\over 10}\dot\tau S_p \approx {2\over 5} ik \Delta_{I\,1} ,
\label{speqn}
\end{equation}
with the solution
\begin{equation}
S_p(k,\eta)\approx {2\over 5}ike^{3\tau(\eta)/10}
\int_0^\eta d\eta'\, e^{-3\tau(\eta')/10}
\Delta_{I\,1}(k,\eta').
\label{spsol}
\end{equation}
Note that the total optical depth $\tau$ is defined as
\begin{equation}
\tau(\eta)=\int_{\eta}^{\eta_\star} \dot\tau(\eta) d\eta,
\label{tau}
\end{equation}
with $\eta_\star$ the conformal time of recombination,
giving $d\tau/d\eta = -\dot\tau$.
We can rewrite the equations for the polarization brightnesses
in a simple form with the change of variables
\begin{eqnarray}
\tilde\Delta_Q &\equiv& e^{ik\mu\eta - \tau}\Delta_Q,\cr
\tilde\Delta_U &\equiv& e^{ik\mu\eta - \tau}\Delta_U,\cr
\tilde S_p &\equiv& e^{ik\mu\eta - \tau} S_p,
\label{tildes}
\end{eqnarray}
giving the evolution equations
\begin{eqnarray}
\dot{\tilde\Delta}_Q(k,\mu,\eta) &=& -{3\over 4}\dot\tau(\eta)
(1-\mu^2)\tilde S_p(k,\eta) + 2\omega_{\bf B}(\mu,\eta)
\tilde\Delta_U(k,\mu,\eta),\cr
\dot{\tilde\Delta}_U(k,\mu,\eta) &=&
-2\omega_{\bf B}(\mu,\eta)
\tilde\Delta_Q(k,\mu,\eta).
\label{tildeeqns}
\end{eqnarray}
We are interested in the case of small Faraday rotation;
the first-order iterative solution to equations~(\ref{tildeeqns}) is
\begin{eqnarray}
\tilde\Delta_Q(k,\mu,\eta) &=& -{3\over 4}(1-\mu^2)
\int_0^\eta d\eta'\dot\tau(\eta') e^{ik\mu\eta'-\tau(\eta')}
S_p(k,\eta'),\cr
\tilde\Delta_U(k,\mu,\eta) &=& {3e^3({\bf B\cdot\hat q})\over
4\pi m^2c\nu^2\sigma_T}(1-\mu^2)
\int_0^\eta d\eta'\dot\tau(\eta')
\int_0^{\eta'} d\eta''\dot\tau(\eta'')
e^{ik\mu\eta''-\tau(\eta'')}S_p(k,\eta'').
\label{solutions}
\end{eqnarray}
Equations (\ref{spsol}) and (\ref{solutions}) 
determine the polarization of
the radiation as a function of frequency, given
the ionization history $x_e$ and the temperature
dipole brightness $\Delta_{I\,1}$. 

Using the polarization brightnesses in $k$-space, we next obtain
an expression for the polarization vector in $x$-space. 
Before performing the Fourier integral, the various
spherical coordinate systems defined for each ${\bf k}$-mode
must be rotated to a common system since the definitions of
the Stokes parameters Q and U depend on the orientation of
the coordinate system (for details, see
Kosowsky 1996). 
The real-space polarization fluctuations are given by
\begin{eqnarray}
{Q({\bf x},\theta,\phi) \over T_0} &=&
{1\over 4}\sum_{\bf k} e^{i{\bf k\cdot x}}
\left[ \Delta_Q({\bf k},\theta',\phi') \cos 2\xi'
+ \Delta_U({\bf k},\theta',\phi') \sin 2\xi' \right],\cr
{U({\bf x},\theta,\phi) \over T_0} &=&
{1\over 4}\sum_{\bf k} e^{i{\bf k\cdot x}}
\left[ -\Delta_Q({\bf k},\theta',\phi') \sin 2\xi'
+ \Delta_U({\bf k},\theta',\phi') \cos 2\xi' \right],
\label{ft}
\end{eqnarray}
where $(\theta',\phi')$ represents the same direction as
$(\theta,\phi)$ except in the coordinate system defined
by the mode $k$. The angle $\xi'$ is the rotation angle to
align the coordinate systems;
an expression for $\xi'$ is given in Kosowsky (1996) but
will not be used in what follows.
Finally, the polarization vector $\bf P$ can be constructed
from these quantities as
\begin{equation}
{\bf P}({\bf x}, \theta,\phi) = {1\over \sqrt{2}}
\left[\hat\theta\sqrt{P(P+Q)} + \hat\phi{U\over |U|}
\sqrt{P(P-Q)}\right],
\end{equation}
with $P\equiv\sqrt{Q^2 + U^2}$.
The angle between two polarization vectors $\varphi_{12}$
follows from
\begin{equation}
\cos^2\varphi_{12} = 
{({\bf P}_1\cdot {\bf P}_2)^2\over P_1^2 P_2^2} =
{1\over 2}\left[1 + {Q_1 Q_2 + U_1 U_2\over P_1 P_2}\right].
\end{equation}

The predictions of a cosmological model are only statistical
in nature. For a given magnetic field, the observationally
interesting quantity is the expectation value
of the Faraday rotation angle. Calculationally, it is useful
to consider only averages of quantities quadratic in
the brightnesses given above; we thus consider the observable
\begin{equation}
\langle \cos^2\varphi_{12}\rangle\approx
{1\over 2}\left[1 + {\left\langle Q_1 Q_2 + U_1 U_2\right\rangle
\over \langle P_1^2\rangle^{1/2}\langle P_2^2\rangle^{1/2} }
\right].
\label{scriptr}
\end{equation}
The averages can be calculated explicitly in terms of
the brightnesses by replacing the averages with the
integral $V^{-1}\int d\bf x$ and replacing the Fourier
sums with integrals, $\sum_{\bf k} \rightarrow [V/(2\pi)^3]
\int d\bf k$, where $V$ is a volume
normalization factor. We also include an exponential beam suppression
to account for a Gaussian beam of width $\sigma$ (Kolb \& Turner 1990). 
The necessary averages can be written
in terms of the two integrals
\begin{eqnarray}
I_Q &\equiv& 
\int_{-1}^1 d\mu \int _0^\infty k^2 dk 
\exp[-4 k^2 H_0^{-2}(1-\mu^2)\sigma^2]
\vert\Delta_Q(k,\mu)\vert^2,\cr
I_U &\equiv& 
\int_{-1}^1 d\mu \int _0^\infty k^2 dk 
\exp[-4 k^2 H_0^{-2}(1-\mu^2)\sigma^2]
|\Delta_U(k,\mu)|^2 
\label{integrals}
\end{eqnarray}
with $H_0$ the Hubble constant; a
long calculation gives the final equations
\begin{eqnarray}
\langle Q_1 Q_2 + U_1 U_2\rangle &=& {V\over 64\pi^2}
(I_Q + f I_U)\cr
\langle P_1^2\rangle &=& {V\over 64\pi^2}
(I_Q + I_U)\cr
\langle P_2^2\rangle &=& {V\over 64\pi^2}
(I_Q + f^2 I_U),
\label{averages}
\end{eqnarray}
where the subscripts 1 and 2 refer to two different
frequencies $\nu_1 < \nu_2$, all of the
brightnesses are evaluated at $\nu_1$, 
and $f\equiv \nu_1^2/\nu_2^2$. 
Since we are comparing
the polarization at two frequencies in the same direction
on the sky, the formulas look like the temperature
correlation function at zero separation.

In the perturbative limit of small rotation angles,
equation~(\ref{scriptr}) reduces to
\begin{equation}
\langle\varphi_{12}^2\rangle^{1/2}
\approx {1\over 2}\left(1-{\nu_1^2\over \nu_2^2}\right)
\left({I_U\over I_Q}\right)^{1/2} .
\label{coslimit}
\end{equation}
Equations~(\ref{solutions}), (\ref{integrals}), 
and (\ref{coslimit}) imply
that the rms rotation angle
is proportional to the magnetic field and
inversely proportional to the square of the frequency,
as expected. An average over all
directions of observation will reduce the above estimate
by a factor of $\sqrt{2}$, due to the changing orientation
of the magnetic field.

\section{Results}

We solve the equations in the previous section numerically
to determine $I_U/I_Q$ and thus the proportionality
factor between the mean
Faraday rotation signal and the quantity $B/\nu^2$. 
Normally, the radiative transfer equations for the
microwave background are expanded in a moment hierarchy 
to eliminate the $\mu$ dependence; however, in this case,
the additional directional dependence of the magnetic
field direction complicates the moment expansion.
We perform the integrals in equations~(\ref{solutions})
on a ($\mu$,$k$) grid. The necessary cosmological
inputs are the dipole component of the radiation intensity,
$\Delta_{I\,1}$, and the differential optical depth through
recombination, $\dot\tau$. 

We choose as an illustrative model ``standard'' cold
dark matter, but the nature of the perturbations has
little effect on our results; the size of the radiation
dipole and the details of recombination affect the final
answer, but neither of these depends strongly on the
cosmological model. The matter density $\Omega_0 h^2$
and the baryon density $\Omega_b h^2$ have a mild effect
on recombination, which we include through numerical
evolution of the free electron density. Hu and Sugiyama (1995a)
give an analytic approximation for $\Delta_{I\,1}$ which we
use here. We evolve the electron density using a
numerical code incorporating the recombination physics detailed
in Hu et al. (1995).

\begin{figure}[htb]
\includegraphics{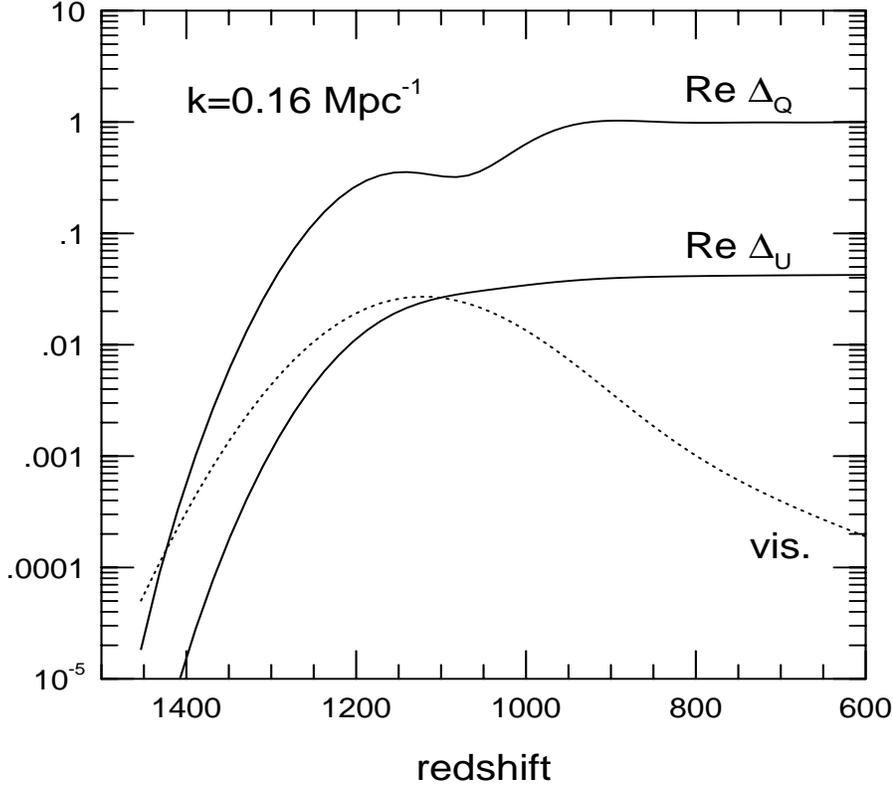}
\vspace*{4.2in}
\caption{The evolution of the polarization brightnesses, for
$k=0.16\,{\rm Mpc}^{-1}$ and $\mu = 0.5$ (in
arbitrary units). Also plotted as
a dotted line is
the differential visibility function $\dot\tau e^{-\tau}$
in units of ${\rm Mpc^{-1}}$.}
\end{figure}

Figure 1 displays the evolution of the polarization 
brightnesses $\Delta_Q$ and $\Delta_U$ through the
last scattering surface for $\mu=0.5$ along
with the differential visibility function $\dot\tau e^{-\tau}$.
At early times, the tight coupling between the photons  
and the baryons prevents the development of polarization. As
decoupling proceeds, the photons begin to free-stream,
generating a quadrupole perturbation which sources
the polarization. However, the induced Faraday rotation
depends on the free electron density, which 
drops to negligible values as recombination ends. Rotation
is generated during the brief period of time when the free electron
density has dropped enough to end tight coupling but not
so much that Faraday rotation ceases. Figure 1 shows that
the generation of rotation lags behind the
generation of polarization, and that rotation is essentially completed
by $z=1100$ whereas polarization generation continues 
almost until $z=900$.

\begin{figure}[htb]
\includegraphics{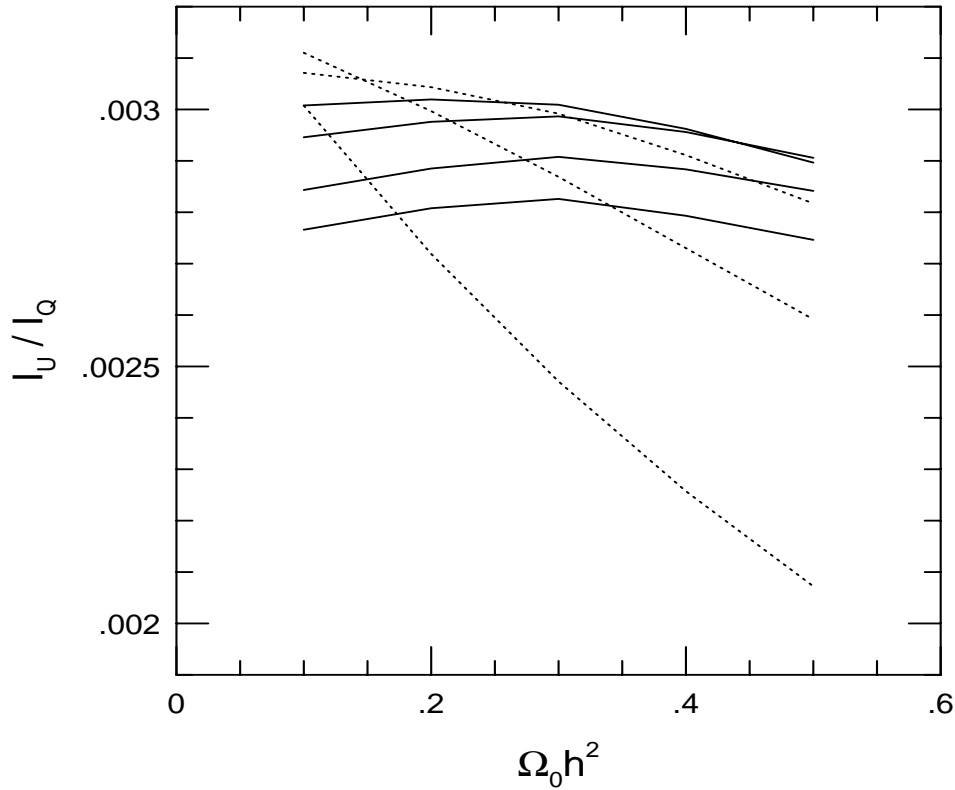}
\vspace*{4.2in}
\caption{The ratio $I_Q/I_U$ as a function of $\Omega_0 h^2$
and $\Omega_b h^2$, for a magnetic field $B_0 = 10^{-9}$ G
and an observed frequency of 30 GHz. The dotted lines from bottom to top
are for $\Omega_b h^2 = 0.005$, 0.0075, and 0.01. The solid
lines from top to bottom are for $\Omega_b h^2 = 0.0125$,
0.015, 0.02, and 0.025.}
\end{figure}

Figure 2 displays $I_U/I_Q$ in equation~(\ref{coslimit}) as a function
of the cosmological parameters $\Omega_0 h^2$, the mass
density parameter, and $\Omega_b h^2$, the baryon density parameter,
for $B_0 = 10^{-9}$ G, $\nu_1 = 30$ GHz, and a beam-width of $0.5^\circ$. 
The beam-width only affects the signal-to-noise ratio of the polarization
signal and has virtually no effect on the size of the rotation angle.
As expected in the Introduction,
the value of $\langle\varphi^2\rangle \propto I_U/I_Q$ is nearly 
independent of these cosmological
parameters. To within a 10\% correction for the effects of
$\Omega_b h^2$ and $\Omega_0 h^2$
(in the range $\Omega_b h^2>0.007$ 
and $\Omega_0h^2<0.3$), the rotation angle is 
\begin{equation}
\langle \varphi_{12}^2\rangle^{1/2} =  1.1^{\circ} \left(1-{\nu_1^2\over
\nu_2^2}\right) 
                  \left({B_0\over 10^{-9}\,{\rm G}}\right)
                  \left({30 \,{\rm GHz}\over \nu_1}\right)^2,
\label{fit}
\end{equation}
\noindent
in good agreement with equation~(\ref{estimate}). We have included here
a factor of $1/\sqrt{2}$ due to an average over all orientations
of the magnetic field.
\vfil\break
 
\section{Discussion}

We have calculated the rotation of the microwave background polarization
vector due to a primordial magnetic field at the surface of last scatter.
The magnetic field was assumed to be coherent on the width of the last
scattering surface, corresponding to a comoving scale of a few Mpc, which
is conveniently the length scale from where galaxies are assembled.  We
have found that the mean Faraday rotation has a size of about $1^\circ$ for
a cosmological magnetic field of $B_0 = 10^{-9}$ G and an observed
frequency of 30 GHz.  An optimized experiment could measure the
orientation of the polarization vector at one frequency which is as low as
practicable, and at a second, somewhat higher frequency. If the frequency
ratio is 2, then 75\% of the total rotation is
observed between the two frequencies.
While extracting this rotation in a single observation
direction is unlikely, a statistical detection averaged over many
observation directions is possible.

The peak amplitude of the expected polarization fluctuations is an
order of magnitude lower than that of the
microwave temperature fluctuations, of order $10^{-6}$
(see, e.g., Crittenden, Davis, \& Steinhardt 1993;
Frewin, Polnarev, \& Coles 1994).  
To measure a Faraday rotation of $1^\circ$ requires another
factor of $10^2$ in sensitivity. The total sensitivity to 
the Faraday rotation signal is proportional to the raw pixel
sensitivity and to the square root of the number of pixels.
Currently envisioned satellite experiments are designed
with the primary goal of mapping the microwave sky temperature down
to small angular scales; given amplifiers of a certain
sensitivity, it is advantageous to design
an experiment with a signal-to-noise ratio per pixel of
around unity and as many pixels on the sky as possible
(Knox 1995). Such a
mapping experiment would require of order $10^6$ pixels to
detect a rotation angle of $1^\circ$. This corresponds to
a beam size of order $10^\prime$ which is at the limit of
current design proposals. It is not unreasonable to
expect that in the coming years, raw sensitivity will
improve so that systematic effects
become the dominant obstacle to detecting the
Faraday rotation signal. As with temperature
measurements, the ultimate limitation will be contamination
from foreground sources. While little is known about
polarization sources at microwave frequencies, estimates
suggest that in the frequency window of 30--80 GHz
the signal should be dominated by the microwave background contribution
(Timbie 1995).
Experiments which focus on high signal-to-noise observations of
small patches of the sky may also prove useful; their feasibility could be 
first demonstrated by searching for a foreground rotation signal 
in the direction of
galaxy clusters with detected rotation measures from radio
source observations
(see, e.g. Dreher et al. 1987; Perley \& Taylor 1991; 
Taylor \& Perley 1993; Ge \& Owen 1993). 
In fact, microwave background polarization measurements 
at multiple frequencies can potentially map the magnetic field
distribution in cluster
environments where the rotation measure is already known to be large
(Loeb \& Kosowsky 1996).

Will foreground contamination of the rotation measure be a considerable
obstacle to measuring the Faraday rotation signal
from primordial recombination? 
If the universe was reionized at a redshift
$z\ll 100\times \Omega^{1/3}(\Omega_b h/0.02)^{-2/3}$,
then the optical depth through reionization is much smaller
than unity and 
the additional Faraday rotation contribution is small.
The contribution from late reionization would be further reduced 
if the coherence length of the primordial field 
is much smaller than a Gpc.
In fact, the cumulative rotation measure of the intergalactic medium 
out to $z\approx 3$ is known to be
smaller than a few ${\rm rad~m^{-2}}$ from samples of high-redshift objects
(see Valee 1990, Kronberg 1994, and references therein).  In addition, 
the rotation measure of the Milky-Way galaxy was measured at high latitudes and
found to be $\lsim 20~{\rm rad~m^{-2}}$ (Spitzer 1978; Simard-Normandin \&
Kronberg 1980). 
According to equation~(\ref{fit}), these foregrounds should allow the
detection of a cosmological field as small as $10^{-10}$ G.

\acknowledgements
We thank George Field for helpful discussions. Wayne Hu
and Naoshi Sugiyama have graciously provided their
code for calculating ionization history through recombination.
This work has been supported in part by the 
Harvard Society of Fellows (for AK) and
NASA ATP grant NAG5-3085 (for AL).

 
\end{document}